\documentclass[11pt]{article}
\pdfoutput=1 
\usepackage{amssymb}
\usepackage{amsmath}
\usepackage{amstext}
\usepackage{graphicx,epsfig}
\usepackage{epsfig}
\usepackage{verbatim} 
\usepackage{fancybox}
\usepackage{color}
\usepackage{ulem}
\usepackage{enumitem}
\usepackage{youngtab}
\usepackage{subfigure}
\usepackage{bbm}
\usepackage{parskip}
\usepackage[numbers,sort&compress]{natbib}
\usepackage{ytableau}
\usepackage[utf8]{inputenc}

\usepackage{tikz}
\usetikzlibrary{positioning, trees,decorations, decorations.markings, decorations.pathmorphing, arrows, shapes.geometric, snakes, patterns}
\pgfdeclaredecoration{complete sines}{initial}
{
    \state{initial}[
        width=+0pt,
        next state=upsine,
        persistent precomputation={\pgfmathsetmacro\matchinglength{
            \pgfdecoratedinputsegmentlength / int(\pgfdecoratedinputsegmentlength/\pgfdecorationsegmentlength)}
            \setlength{\pgfdecorationsegmentlength}{\matchinglength pt}
        }] {}
    \state{upsine}[width=\pgfdecorationsegmentlength,next state=downsine]{
        \pgfpathsine{\pgfpoint{0.25\pgfdecorationsegmentlength}{0.5\pgfdecorationsegmentamplitude}}
        \pgfpathcosine{\pgfpoint{0.25\pgfdecorationsegmentlength}{-0.5\pgfdecorationsegmentamplitude}}
    }
    \state{downsine}[width=\pgfdecorationsegmentlength,next state=upsine]{
        \pgfpathsine{\pgfpoint{0.25\pgfdecorationsegmentlength}{-0.5\pgfdecorationsegmentamplitude}}
        \pgfpathcosine{\pgfpoint{0.25\pgfdecorationsegmentlength}{0.5\pgfdecorationsegmentamplitude}}
}
    \state{final}{}
}

\newcommand{\Comment}[1]{{}}
\definecolor{darkblue}{rgb}{0.15,0.35,0.55}
\definecolor{reddish}{rgb}{0.65, 0.2, 0.2}
\usepackage[linktocpage=true]{hyperref}
\hypersetup{
colorlinks=true,
citecolor=darkblue,
linkcolor=reddish,
urlcolor=darkblue,
pdfauthor={},
pdftitle={},
pdfsubject={}
}

\definecolor{green3}{RGB}{44, 160, 44}

\newcommand{\be}{\begin{equation}}
\newcommand{\ee}{\end{equation}}
\newcommand{\bea}{\begin{eqnarray}}
\newcommand{\eea}{\end{eqnarray}}
\newcommand{\beas}{\begin{eqnarray*}}
\newcommand{\eeas}{\end{eqnarray*}}
\newcommand{\nn}{\nonumber}

\definecolor{darkred}{rgb}{0.7,0.3,0.3}
\definecolor{darkgreen}{rgb}{0.2,0.7,0.3}
\definecolor{lightgreen}{rgb}{.816,.94,.753}
\definecolor{greyish}{rgb}{.8,.8,.8}
\definecolor{darkblue2}{rgb}{0.3,0.4,0.9}
\usepackage{pifont}
\usepackage{xcolor,colortbl}

\def\({\left(}
\def\){\right)}

\newcommand{\ra}{\rangle}

\def\gsim{ \lower .75ex \hbox{$\sim$} \llap{\raise .27ex \hbox{$>$}} }
\def\lsim{ \lower .75ex \hbox{$\sim$} \llap{\raise .27ex \hbox{$<$}} }

\def\xyma{\xymatrix@M.7em}
\def\xymas{\xymatrix@M.1em}

\title{}
\author{}

\numberwithin{equation}{section}

\begin{document}
\tikzset{
    photon/.style={decorate, decoration={snake}, draw=magenta},
    graviton/.style={decorate, decoration={snake}, draw=black},
 sgal/.style={decorate , dashed, draw=black},
 scalar/.style={decorate , draw=black},
  mgraviton/.style={decorate, draw=black,
    decoration={coil,amplitude=4.5pt, segment length=7pt}}
    electron/.style={draw=blue, postaction={decorate},
        decoration={markings,mark=at position .55 with {\arrow[draw=blue]{>}}}},
    gluon/.style={decorate, draw=magenta,
        decoration={coil,amplitude=4pt, segment length=5pt}} 
}
\renewcommand{\thefootnote}{\fnsymbol{footnote}}
~

\begin{center}
{\huge \bf On the Conformal Symmetry of \\ Exceptional Scalar Theories\\
}
\end{center} 

\vspace{1truecm}
\thispagestyle{empty}
\centerline{\Large Kara Farnsworth,${}^{\rm a,}$\footnote{\href{mailto:kmfarnsworth@gmail.com}{\texttt{kmfarnsworth@gmail.com}}} Kurt Hinterbichler,${}^{\rm a,}$\footnote{\href{mailto:kurt.hinterbichler@case.edu} {\texttt{kurt.hinterbichler@case.edu}}} Ond\v{r}ej Hul\'{i}k,${}^{\rm a,b,}$\footnote{\href{mailto:ondra.hulik@gmail.com} {\texttt{ondra.hulik@gmail.com}}} }

\vspace{.5cm}

\centerline{{\it ${}^{\rm a}$CERCA, Department of Physics,}}
 \centerline{{\it Case Western Reserve University, 10900 Euclid Ave, Cleveland, OH 44106}} 
 \vspace{.25cm}
 
 \centerline{{\it ${}^{\rm b}$Institute of Physics of the Czech Academy of Sciences, CEICO,}}
 \centerline{{\it Na Slovance 2, 182 21 Prague 8, Czech Republic}} 

 \vspace{1cm}
\begin{abstract}
\noindent

The DBI and special galileon theories exhibit a conformal symmetry at unphysical values of the spacetime dimension.  We find the Lagrangian form of this symmetry.  The special conformal transformations are non-linearly realized on the fields, even though conformal symmetry is unbroken. Commuting the conformal transformations with the extended shift symmetries, we find new symmetries, which when taken together with the conformal and shift symmetries close into a larger algebra.  For DBI this larger algebra is the conformal algebra of the higher dimensional bulk in the brane embedding view of DBI.  For the special galileon it is a real form of the special linear algebra.  We also find the Weyl transformations corresponding to the conformal symmetries, as well as the necessary improvement terms to make the theories Weyl invariant, to second order in the coupling in the DBI case and to lowest order in the coupling in the special galileon case.

\end{abstract}

\newpage

\setcounter{tocdepth}{2}
\tableofcontents
\renewcommand*{\thefootnote}{\arabic{footnote}}
\setcounter{footnote}{0}

\newpage

\section{Introduction}

Soft theories are effective field theories whose amplitudes go to zero as one of the external momenta are scaled to zero, scaling with some power of this momentum as we approach zero.  Among soft theories consisting of a single scalar $\phi$, there are two exceptional theories that scale with the highest possible power, a power higher than expected given the number of derivatives per field in the Lagrangian \cite{Cheung:2014dqa,Cheung:2016drk}.  These are the Dirac-Born-Infeld (DBI) theory, whose action is given in \eqref{lageeqflat}, and the special galileon \cite{Cheung:2014dqa,Hinterbichler:2015pqa}, whose action is given in \eqref{eqsgal}.  Both theories depend on a single dimensionful coupling constant $\alpha$, and are otherwise completely fixed. 

These two theories, among others \cite{Carrasco:2019qwr}, have a privileged place among scalar effective field theories; they naturally appear in the Cachazo-Yuan-He (CHY) representation \cite{Cachazo:2013hca, Cachazo:2014xea,Cachazo:2016njl} and act as nodes in a web of theories related by the double copy and other procedures \cite{Cheung:2017ems,Cheung:2017yef} (see \cite{Carrillo-Gonzalez:2018pjk} for a nice visual representation of this web). Understanding their properties and what makes these theories unique is a step towards being able to uncover how fundamental these tools are and their ability to relate gauge and gravitational theories.

DBI and the special galileon exist for any spacetime dimension $d$, however in \cite{Cheung:2020qxc} it was noted that if we consider these theories at unphysical values of $d$, $d=0$ for DBI and $d=-2$ for the special galileon, then the coupling $\alpha$ becomes dimensionless and the theory is scale invariant.  In fact, each term in the Lagrangian is separately scale invariant, as classical scale invariance is automatic and trivial once there are no dimensionful couplings.  In \cite{Cheung:2020qxc} it is shown that these theories also have full conformal invariance in addition to scale invariance, and that the conformal invariance is non-trivial and fixes the structure of the Lagrangian to be that of DBI/special galileon.  For example, it was argued that the stress tensor on flat space could be improved to be traceless, and that the amplitudes satisfy conformal Ward identities, only if the structure of the theory was that of DBI/special galileon.

However, conformal symmetry as we usually understand it is linearly realized on the fields, and a linear symmetry can never relate terms with different powers of the field, as would be required to fix the full non-linear structure of these theories.  We therefore expect that this conformal symmetry is realized in a novel non-linear way.  On the other hand, conformal symmetry should still be unbroken, so that the usual Ward identities on the amplitudes are satisfied.  Here we resolve this puzzle and find the field transformations that leave the Lagrangians invariant and fix their non-linear structure, without spontaneously breaking conformal symmetry.  The transformations include non-linearities in the special conformal generators (shown in \eqref{conformalffieldrepsenew} for DBI and \eqref{conformalffieldrepsenewge} for the special galileon), but preserve the $\phi=0$ vacuum and satisfy the same conformal algebra as their linear counterparts.

Another defining feature of these theories is that they possess non-trivial shift symmetries, where the field shifts by powers of the spacetime coordinate $x^\mu$.  The shift symmetries for DBI have up to one power of $x^\mu$ (shown in \eqref{shiftsymmesdbiee}), and the shift symmetries for the special galileon have up to two powers of $x^\mu$ (shown in \eqref{galshifts}).  These shift symmetries also fix the full non-linear structure of the theory, and are responsible for the enhanced soft limits.  

These shift symmetries do not close with the conformal symmetries under commutation, and instead produce new symmetries.  In the DBI case, we find a new symmetry \eqref{newsymmt} which has 2 powers of $x^\mu$, and in the special galileon  case we find several new symmetries \eqref{newtransgalee} which have up to 4 powers of $x^\mu$.  These new symmetries combined with the conformal symmetries and shift symmetries then close to form a larger algebra.  In the DBI case, this larger algebra is the conformal algebra of one dimension higher.  In Section \ref{DBIbranesection}, we explain this from the brane embedding point of view; the DBI theory can be thought of as the worldvolume theory of a $d$ dimensional brane embedded in a $d+1$ dimensional bulk, and from this perspective, the symmetries all descend from conformal symmetries of the bulk.  In the special galileon case, the full symmetry algebra is a real form of the special linear algebra.

Finally, we consider coupling these theories to a metric and look for a local Weyl symmetry that descends to the new non-linear global conformal symmetries when the metric is frozen to flat space.  This Weyl symmetry requires new non-linear terms in the scalar  field transformation in order to reproduce the non-linear terms in the global special conformal transformations.  We find the improvement terms necessary to add to the theories so that they are Weyl invariant, to lowest order in the coupling for the special galileon, and to second order in the coupling for DBI. 

\paragraph{Conventions:} The spacetime dimension is $d$.  We use the mostly plus metric.   The curvature conventions are those of \cite{Carroll:2004st}.

\section{DBI}

The DBI action for a single scalar $\phi$ in $d$ spacetime dimensions is
\be { S}=\int d^dx\  -{1\over \alpha }\sqrt{1+\alpha (\partial\phi)^2}\, .\label{lageeqflat}\ee
It depends on a single coupling constant, $\alpha$.  

In any $d$, this action is manifestly invariant under the standard translations and Lorentz transformations,
\bea P^\mu \, \phi = -\partial^\mu \phi  ,\ \ \ \ 
J^{\mu\nu} \, \phi = (x^\mu\partial^\nu-x^\nu\partial^\mu)\phi  \,, \label{pooincaresymse}
\eea
which satisfy the commutation relations of the Poincare algebra,
\bea \left[J^{\mu\nu},J^{\sigma\rho}\right]&=&\eta^{\mu\sigma}J^{\nu\rho}-\eta^{\nu\sigma}J^{\mu\rho}+\eta^{\nu\rho}J^{\mu\sigma}-\eta^{\mu\rho}J^{\nu\sigma} \,,\nn\\
\left[J^{\mu\nu},P^\rho\right] &=& \eta^{\rho\mu} P^\nu-\eta^{\rho\nu} P^\mu\, ,\nn \\
\left[P^\mu,P^\nu\right]&=&0 \,.\label{poincaecomese}
\eea

\subsection{Conformal symmetry\label{conformaldbis}}

The DBI coupling constant in \eqref{lageeqflat} has mass dimension $[\alpha]=-d$.  As was noted in \cite{Cheung:2020qxc}, if we consider the unphysical value $d=0$, the coupling becomes dimensionless, and the theory becomes scale invariant under the standard dilation symmetry $D\phi= -\left(x^\mu\partial_\mu+\Delta\right)\phi$, where $\Delta$ is the scaling dimension of the field.  Scale invariance requires the field $\phi$ to have dimension $\Delta={d-2\over 2}$, so when $d=0$ we have $\Delta=-1$, and the action \eqref{lageeqflat} is invariant under \be D\phi= -x^\mu\partial_\mu\phi+\phi \,. \label{scalesyme}\ee
This scale symmetry, like the Poincare transformations \eqref{pooincaresymse}, is linear in the fields, and each term in the expansion of the square root in the action \eqref{lageeqflat} in powers of $\alpha$ is separately scale invariant when $d=0$\footnote{In checking invariance for this and other symmetries below in $d=0$, we proceed as in dimensional regularization, manipulating everything in general $d$, setting $d=0$ only at the end and never using any dimensionally dependent identities.}.

Conformal symmetry also includes special conformal transformations, and the standard linear action of these on a field of weight $\Delta$ is 
\be K^\mu\, \phi =\left(-2x^\mu x^\nu\partial_\nu+x^2 \partial^\mu -2 x^\mu\Delta\right)\phi \label{linconfsyme}.\ee
 In \cite{Cheung:2020qxc} it was argued that conformal symmetry should fix the square root structure in the DBI action when $d=0$; a generic scalar effective theory in $d = 0$ with one derivative per field would be scale invariant, but not conformally invariant, and only the particular square root structure of DBI would have full conformal invariance.   However this cannot be achieved with the standard linear special conformal transformations  \eqref{linconfsyme}, since they cannot relate terms with differing powers of $\phi$ in order to fix the square root structure. 
 
 The non-linear special conformal transformation which does accomplish this is
\be 
K^\mu\, \phi =\left(-2x^\mu x^\nu\partial_\nu+x^2 \partial^\mu +2 x^\mu\right)\phi +\alpha \phi^2\partial^\mu\phi. \label{conformalffieldrepsenew}
\ee
This transformation contains a new non-linear piece which depends on $\alpha$, and when $\alpha=0$ it reduces to the standard linear special conformal transformation \eqref{linconfsyme} with $\Delta=-1$.  The non-linear transformation \eqref{conformalffieldrepsenew} is a symmetry of the DBI action when $d=0$, and it fixes the square root structure of the action.  Note that despite the non-linearities, the transformation \eqref{conformalffieldrepsenew} preserves the vacuum $\phi=0$, so we can still say that conformal symmetry is unbroken\footnote{A special conformal transformation similar to \eqref{conformalffieldrepsenew} occurs in the worldvolume theory of a flat brane probing an AdS bulk, but in this case conformal symmetry is broken \cite{Hinterbichler:2012fr}.}.  Usually non-linear transformations are associated with broken symmetries, but in these cases there is always a leading term in the transformation that is independent of the fields so that the vacuum is not preserved.  Here we have no such term, the transformation starts at linear order in $\phi$.

We can now compute the commutators of the scale symmetry \eqref{scalesyme} and non-linear special conformal symmetries \eqref{conformalffieldrepsenew} with the Poincare symmetries \eqref{pooincaresymse} and we get
\bea \left[P^\mu,D\right]&=& P^\mu\,, \nn\\
 \left[K^\mu,D\right]&=& -K^\mu\,, \nn\\
  \left[K^\mu,P^\nu\right]&=&-2(\eta^{\mu\nu} D-J^{\mu\nu})\,,\nn\\
   \left[J^{\mu\nu},K^\rho\right]&=& \eta^{\rho\mu}K^\nu-\eta^{\rho\nu}K^\mu, \nn\\
   \left[K^\mu,K^\nu\right]&=&\left[J^{\mu\nu},D\right]=0\,, \label{conformalcommutatorese}
   \eea
along with the Poincare commutators \eqref{poincaecomese}.  These are the standard commutation relations of the conformal algebra.  They hold independently of $\alpha$, so despite the non-linear term in \eqref{conformalffieldrepsenew}, we have the usual conformal algebra.  In computing these we have not fixed the spacetime dimension $d$, though we did need to fix $\Delta=-1$ in writing the generators \eqref{scalesyme}, \eqref{conformalffieldrepsenew}.  As is well known, we can define
$ J^{-1,0}=D,\ J^{-1,\mu}={1\over 2}(P^\mu-K^\mu),\  J^{0,\mu}={1\over 2}(P^\mu+K^\mu),$
and assemble the conformal generators into a $d+2$ dimensional  anti-symmetric matrix $J^{AB}$ with $A,B=-1,0,1,\cdots,d$
\be J^{AB}\equiv\left(\begin{array}{c|c|c}0 & D &  {1\over 2}(P^\nu-K^\nu) \\\hline -D & 0 & {1\over 2}(P^\nu+K^\nu) \\\hline -{1\over 2}(P^\mu-K^\mu) & -{1\over 2}(P^\mu+K^\mu) & J^{\mu\nu}\end{array}\right).\label{Jmatrixgenembsee}\ee
Then the commutation relations \eqref{conformalcommutatorese}, \eqref{poincaecomese} become 
\be \left[J^{AB},J^{CD}\right]=g^{AC}J^{BD}-g^{BC}J^{AD}+g^{BD}J^{AC}-g^{AD}J^{BC}\, ,\label{Jcommsd2e}\ee
where
\be g^{AB}\equiv\left(\begin{array}{c|c|c}-1 &   &   \\\hline   & 1 &   \\\hline   &   & \eta^{\mu\nu}\end{array}\right)\,,\label{D2metricdefee}\ee
showing that the conformal algebra is $so(2,d)$.

Since the field $\phi$ has conformal dimension $\Delta=-1$ in dimension $d=0$, it might be thought that it satisfies the standard scalar unitarity bound $\Delta\geq {d-2\over 2}$ \cite{Mack:1975je,Jantzen1977}.   This unitarity bound comes from demanding positivity of the second level descendent state $P^2|\psi\ra$, given that the primary $|\psi\ra$ has positive norm.  But there is also a bound $\Delta\geq 0$ coming from positivity of the first level descendant $P^\mu|\psi\ra$.  When $d\geq 2$ this bound is subsumed by the standard bound, but in our case where $d=0$, $\Delta=-1$ it is stronger and it is violated.  This is a moot point, however, since the non-linear action \eqref{conformalffieldrepsenew} of the conformal algebra is not the standard one that leads to these bounds, and so the bounds may not apply.

\subsection{Including shift symmetries}

The DBI action \eqref{lageeqflat} is also invariant in any $d$ under the extended shift symmetries,
\bea C \, \phi = 1 ,\ \ \ 
B^\mu\, \phi =x^\mu+\alpha\, \phi\partial^\mu\phi , \label{shiftsymmesdbiee}
\eea
where $x^\mu$ is the spacetime coordinate.
These are spontaneously broken, since they do not preserve the vacuum $\phi=0$.  Any theory with one derivative per field has the simple shift symmetry $C$, but the extended shifts $B^\mu$, which include terms linear in the spacetime coordinate $x^\mu$, are $\alpha$ dependent and fix the square root structure of the action in any $d$.  

We can now ask how these shift symmetries interplay with the conformal symmetries \eqref{pooincaresymse}, \eqref{scalesyme}, \eqref{conformalffieldrepsenew} by computing the commutators.  In fact, taken together, the conformal symmetries and shift symmetries \eqref{shiftsymmesdbiee} do not quite close under the commutators.  By computing $[K^\mu,B^\nu]$, we find a new scalar symmetry,
\be N\phi=-x^2+\alpha \left(\phi^2-2 x^\mu \phi \partial_\mu\phi \right).\label{newsymmt}\ee
This new transformation is also a symmetry of DBI when $d=0$, and it starts with a quadratic shift in the spacetime coordinate. (For example, we can quickly see that the kinetic term $\sim(\partial\phi)^2$ is invariant under the lowest order shift $N^{(0)}\phi\sim x^2$: $N^{(0)}\left((\partial\phi)^2)\right)\sim \partial_\mu\phi \partial^\mu( N^{(0)}\phi) \sim \partial_\mu\phi x^\mu$, which integrates by parts to $\sim d \phi$, which vanishes when $d=0$.)

After including $N$, the algebra closes.  The complete set of commutators, in addition to the conformal algebra \eqref{conformalcommutatorese}, \eqref{poincaecomese}, is
\bea  && \left[J^{\mu\nu},B^\rho\right] = \eta^{\rho\mu} B^\nu-\eta^{\rho\nu} B^\mu \,,\ \ \ \ 
  \left[J^{\mu\nu},C\right] =   \left[J^{\mu\nu},N\right] = 0 \,, \label{Dbishift1} \\
&&   \left[N,D \right]=  -N\,,\ \ \ 
  \left[B^\nu,D\right]=0\, ,\ \ \ 
    \left[C,D \right]= C \,,  \label{Dbishift2} \\
  &&  \left[P^\mu,N  \right]=-2B^\mu \, ,\ \ \   \left[P^\mu,B^\nu\right]=\eta^{\mu\nu} C\,, \ \ \ \left[P^\mu,C\right]=0\,, \label{Dbishift3}  \\
  &&  \left[K^\mu ,C \right]= -2B^\mu\,,\ \ \  \left[K^\mu,B^\nu\right]=\eta^{\mu\nu} N\, ,\ \ \   \left[K^\mu ,N \right]=0 \,, \label{Dbishift4} 
  \eea
\bea &&  \left[B^\mu,B^\nu\right]=\alpha J^{\mu\nu} ,\ \ \ \left[C,B^\mu\right]= -\alpha P^\mu\, ,\ \ \  \left[N,B^\mu\right]=-\alpha K^\mu\, ,\ \ \  \left[C,N \right]=  2\alpha D \,.  \label{Dbishift5} 
\eea
As before, we do not need to fix $d=0$ in computing these, though we have used $\Delta=-1$ in defining the generators.  

The relations \eqref{Dbishift1} indicate that the shift symmetries $N,B^\mu,C$ transform under Lorentz transformations simply as their tensor indices indicate, with spins $0, 1, 0$ respectively.  The relations \eqref{Dbishift2} indicate that they are eigenvectors under the action of commutation with $D$ to the right, with eigenvalues $-1,0,1$ respectively.  The relations \eqref{Dbishift3} and \eqref{Dbishift4} indicate that $P^\mu$ and $K^\mu$ are like raising and lowering operators respectively between the different $D$ eigenstates.  These relations  \eqref{Dbishift1}, \eqref{Dbishift2}, \eqref{Dbishift3}, \eqref{Dbishift4} are all the commutators between elements of the conformal algebra and shift symmetries, and are independent of $\alpha$.  They can be summarized in this figure:
\be
 \ \ \ \ \ \raisebox{-96pt}{\epsfig{file=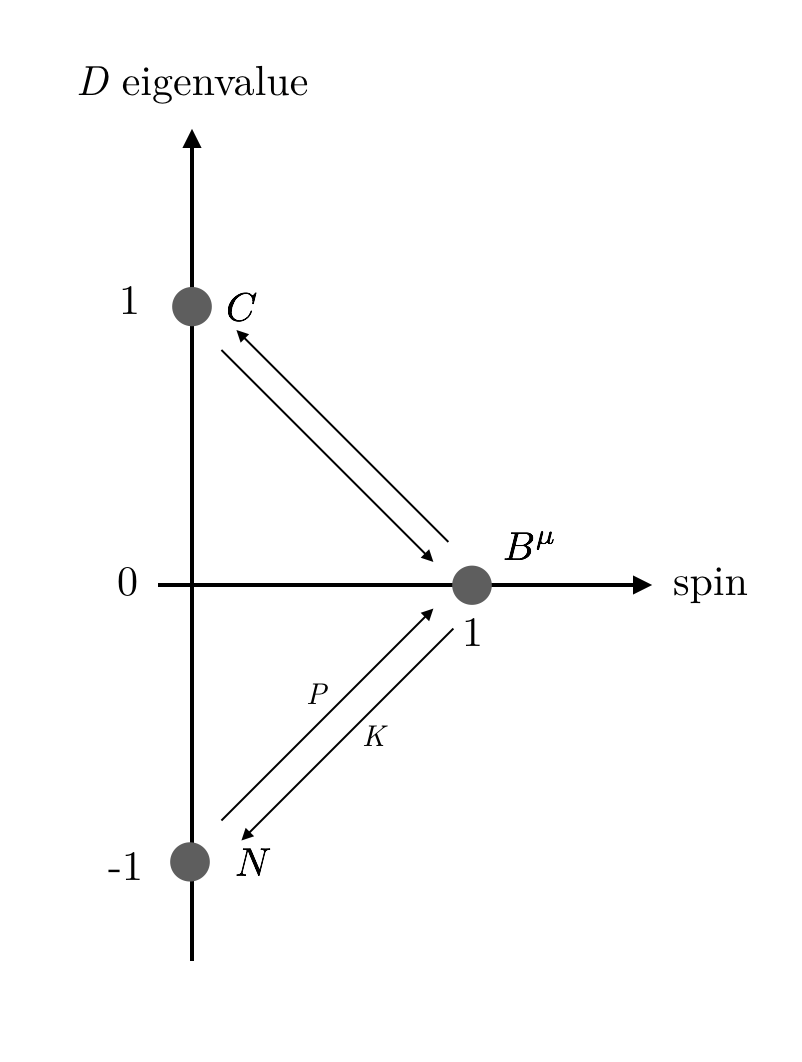,height=2.7in}}\label{DBImodule}
\ee
The DBI symmetries thus form a finite dimensional Verma module under the conformal 
algebra, where the new symmetry $N$ is like a conformal primary and $B^\mu$, $C$ are descendants.  This short finite dimensional module, which occurs when a scalar conformal primary has dimension $-1$ as is the case here\footnote{This is a type I shortening condition in the language of \cite{Penedones:2015aga}, which occur for scalar primaries when they have weight $0,-1,-2,\cdots$.  The trivial case with weight 0 would describe a simple shift symmetric field with conformal symmetry.  The next case with weight $-2$ will occur in Section \ref{sgalshiftsec} for the special galileon. }, is nothing but the fundamental vector representation of the $so(2,d)$ conformal algebra.  To see this, define  $S^{-1}=-{1\over 2}\left(C-N\right),\  S^{0}=-{1\over 2}\left(C+N\right),\ S^\mu=B^\mu,$ and assemble the DBI symmetries into a $d+2$ dimensional vector $S^A$, 
\be S^A=\left(\begin{array}{c}-{1\over 2}\left(C-N\right) \\\hline -{1\over 2}\left(C+N\right)  \\\hline B^\mu \end{array}\right)\,.
\ee
Then the commutators \eqref{Dbishift1}, \eqref{Dbishift2}, \eqref{Dbishift3}, \eqref{Dbishift4} become
\be  \left[J^{AB},S^C\right] = g^{AC} S^B-g^{CB} S^A,\ \ \label{SJcommse}\ee
with the conformal generators packaged into $J^{AB}$ as in \eqref{Jmatrixgenembsee} with $g^{AB}$ as in \eqref{D2metricdefee}.
This is the statement that $S^A$ is a vector under the conformal algebra $so(2,d)$.
The relations \eqref{Dbishift5}, which give the commutators between the shift symmetries, are proportional to $\alpha$, and can now be summarized as
\be  \left[S^A,S^B\right]=\alpha J^{AB}\,.\label{SScommse}\ee
  
To identify the full algebra, we assemble $J^{AB}$ and $S^A$ into an $d+3$ dimensional anti-symmetric matrix $J^{{\cal A}{\cal B}}$ with ${\cal A},{\cal B}=-1,0,1,\cdots,d,d+1$,
\be J^{{\cal A}{\cal B}}=\left(\begin{array}{cc} J^{AB} & -{1\over \sqrt{\alpha}}S^B \\ {1\over \sqrt{\alpha}}S^A & 0\end{array}\right)\, , \ \ \alpha>0\,. \label{refcomupdJeqe}\ee
The commutators \eqref{SScommse}, \eqref{SJcommse}, \eqref{Jcommsd2e} can now all be summarized as
\be \left[J^{{\cal A}{\cal B}},J^{{\cal C}{\cal D}}\right]=g^{{\cal A}{\cal C}}J^{{\cal B}{\cal D}}-g^{{\cal B}{\cal C}}J^{{\cal A}{\cal D}}+g^{{\cal B}{\cal D}}J^{{\cal A}{\cal C}}-g^{{\cal A}{\cal D}}J^{{\cal B}{\cal C}} ,\ee
where
\be g^{{\cal A}{\cal B}}=\left(\begin{array}{cc}g^{AB} & 0 \\0 & 1\end{array}\right),\ \ \ \alpha>0 \,.\label{D3metrice}\ee
These are the commutation relations of $so(2,d+1)$.  In writing \eqref{refcomupdJeqe}, \eqref{D3metrice}, we have assumed that $\alpha>0$, which is the ``right sign"  from the standard UV completion point of view \cite{Adams:2006sv}.  If instead we have the ``wrong sign'' $\alpha<0$ \cite{Cooper:2013ffa}, then the $d+3$ metric \eqref{D3metrice} will have a minus sign in the lower right corner, and the full algebra will be $so(3,d)$.

The smallest symmetry algebra of DBI that includes the conformal symmetry of \cite{Cheung:2020qxc} and the shift symmetries is thus isomorphic to the conformal algebra of a space of dimension $d+1$.  The breaking pattern is to the $d$ dimensional conformal algebra, since the shifts \eqref{shiftsymmesdbiee}, \eqref{newsymmt} are the only ones that do not preserve the vacuum $\phi=0$,
\be so(2,d+1)\rightarrow so(2,d)\,,\ \ \ \alpha>0.\ee
In fact, this is isomorphic to the symmetry breaking pattern of the AdS DBI theory, the theory of a $d+1$ dimensional AdS brane in a $d+2$ dimensional AdS bulk \cite{Goon:2011qf,Goon:2011uw} (or the $k=1$ non-abelian AdS scalar shift theory, in the language of \cite{Bonifacio:2018zex}).

\subsection{Brane point of view\label{DBIbranesection}}

The DBI theory \eqref{lageeqflat} can be thought of as a gauge fixed worldvolume theory of a $d$ dimensional flat brane probing a $d+1$ dimensional flat bulk \cite{deRham:2010eu}.
The fact that the full symmetry algebra including the conformal and shift symmetries is the conformal algebra of one dimension higher suggests that they have their origin as conformal symmetries of the bulk.  Here we will see that this is indeed the case. 
(In this subsection we'll assume $\alpha>0$ and suppress the coupling, for clarity.)

As detailed in \cite{Hinterbichler:2010xn,Goon:2011qf}, we consider the theory of scalars $X^A(x)$, $A=1,\cdots,d+1$, describing the embedding of a $d$ dimensional worldvolume into a $d+1$ dimensional bulk.  The fixed bulk metric is $G_{AB}(X)$, and the 
 induced metric on the brane is
\be \bar g_{\mu\nu}(x)={\partial X^A\over \partial x^\mu}{\partial X^B\over\partial x^\nu}G_{AB}(X(x)).\label{inducedmetge}\ee
The action is
\be S = -\int d^dx\ \sqrt{-\bar g},\label{ungaguefixee}\ee
and is manifestly invariant under worldvolume diffeomorphisms $\xi^\mu(x)$, under which the $X^A$ transform as scalars $\delta_\xi X^A=\xi^\mu\partial_\mu X^A$.
 
Given a transformation by a bulk diffeomorphism $K^A(X)$, $\delta X^A=K^A(X)$,  the induced metric is invariant if $K^A$ is a Killing vector of the bulk metric, 
\be
{\cal L}_K G_{AB}\equiv K^C\partial_C G_{AB}+\partial_AK^CG_{CB}+\partial_BK^CG_{AC}=0,
\ee
and this is a global symmetry of the action in any $d$.

Suppose instead that $K^A$ is a conformal Killing vector, 
\be {\cal L}_K G_{AB}=\Phi G_{AB},\ee
where $\Phi={2\over d+1}\nabla_A  K^A$.  Then under such a diffeomorphism we have
\be \delta \bar g_{\mu\nu}=\Phi \bar g_{\mu\nu},\ee
and the change in the action \eqref{ungaguefixee} is then
\be \delta S = - \int d^dx\  \delta\left( \sqrt{-\bar g}\right)=  -{d\over 2}\int d^dx\ \sqrt{-\bar g} \Phi\,.  \ee
This vanishes when $d=0$, which means that the shift by a bulk conformal Killing vector is a symmetry of \eqref{ungaguefixee} when $d=0$.

To recover the DBI theory \eqref{lageeqflat}, we consider a flat bulk metric $G_{AB}(X)=\eta_{AB}$, and we fix the worldvolume diffeomorphisms by going to the unitary gauge $X^\mu=x^\mu$, $X^{d+1}=\phi$.  The induced metric \eqref{inducedmetge} becomes $\bar g_{\mu\nu}=\eta_{\mu\nu}+\partial_\mu\phi\partial_\nu\phi$, and the action \eqref{ungaguefixee} becomes the DBI action \eqref{lageeqflat}.
The transformation resulting from a bulk diffeomorphism $K^A$ must be compensated by a worldvolume diffeomorphism with $\xi^\mu=-K^\mu$ in order to preserve the unitary gauge condition, so that the resulting global transformation acting on $\phi$ is
\be \delta_K\phi=-K^\mu(x,\phi)\partial_\mu\phi+K^{d+1}(x,\phi).\label{Ktounitaryge}\ee
When the $K^A$ are isometries of $\eta_{AB}$, we get global symmetries valid in any $d$; the isometries along the brane directions $\mu$ give the Poincare symmetries \eqref{pooincaresymse}, the translation in the $d+1$ direction gives the shift $C$ in \eqref{shiftsymmesdbiee}, and the Lorentz boosts mixing the $d+1$ and $\mu$ directions give the extended shifts $B^\mu$ in \eqref{shiftsymmesdbiee}.

We can now ask about the conformal Killing vectors of $\eta_{AB}$ that are not Killing vectors, which will give symmetries only when $d=0$.
The bulk dilation vector field is
\be K^A(X)=X^A,\ee
which plugging into \eqref{Ktounitaryge} leads to
\be \delta\phi=-x^\mu\partial_\mu\phi+\phi,\ee
which is precisely the dilation in \eqref{scalesyme}, including the correct scaling weight $\Delta=-1$.
The bulk special conformal vector fields are
\be K^A_{(B)}(X)=2X_BX^A-X^2\delta^A_B.\ee
Looking at the $B=\mu$ components, plugging into \eqref{Ktounitaryge} leads to
\be \delta_{(\mu)}\phi=\left(-2x^\mu x^\nu\partial_\nu+x^2 \partial^\mu +2 x^\mu\right)\phi + \phi^2\partial^\mu\phi,\ee
which, once the coupling is restored, is precisely the special conformal transformation in \eqref{conformalffieldrepsenew}, with the correct scaling weight $\Delta=-1$, and including the new non-linear term.
Looking at the $B=d+1$ component, plugging into \eqref{Ktounitaryge} leads to
\be \delta_{(d+1)}\phi=-x^2+ \left(  \phi^2-2 x^\mu \phi \partial_\mu\phi\right)\,,\ee
which, once the coupling is restored, is precisely the new transformation in \eqref{newsymmt}.

This construction makes it clear that new symmetries for $d=0$ will also be present in any brane theory where the bulk has conformal Killing vectors which are not Killing vectors.  This includes the AdS and dS DBI theories discussed in \cite{Goon:2011qf,Goon:2011uw,Burrage:2011bt,Trodden:2011xh,Bonifacio:2018zex}, and more general cases such as the cosmological setup in \cite{Goon:2011xf}.

\section{Special Galileon}

We now turn to the special galileon.  The action can be written as
\bea
S_{\rm sgal} &&=-\frac{1}{2}\int d^dx \sum_{n=1}^{ \left \lfloor \frac{d+1}{2} \right \rfloor}  \frac{\alpha^{n-1}}{(2n-1)!} \left( \partial \phi \right)^2 \mathcal{L}^{\rm TD}_{2n-2} \nn\\
&&= \int d^dx -\frac{1}{2}(\partial\phi)^2 - \frac{\alpha}{12}(\partial\phi)^2\Big[(\square\phi)^2 - (\partial_\mu\partial_\nu\phi)^2\Big]+\cdots\,, \label{eqsgal}
\eea
where the total derivative combinations are $\mathcal{L}^{\rm TD}_{n} \equiv \sum_p (-1)^p \eta^{\mu_1 p({\nu_1})} \cdots \eta^{\mu_n p({\nu_n})} \partial_{\mu_1 }\partial_{\nu_1}\phi \cdots \partial_{\mu_n}\partial_{ \nu_n}\phi$
with the sum running over all permutations of the $\nu$ indices with $(-1)^p$ the sign of the permutation.
As with DBI, the theory depends on a single coupling constant, $\alpha$, and in any $d$ it is manifestly invariant under the standard translations and Lorentz transformations \eqref{pooincaresymse}.

\subsection{Conformal symmetry\label{sgalconfsection}}

The galileon coupling constant in \eqref{eqsgal} has dimension $[\alpha]=-(d+2)$.  As was noted in \cite{Cheung:2020qxc}, if we consider the unphysical value $d=-2$, then the coupling becomes dimensionless, and the theory becomes scale invariant.    In this case the theory will have the standard dilation symmetry $D\phi= -\left(x^\mu\partial_\mu+\Delta\right)\phi$, where $\Delta$ is the conformal dimension of the field.  The field
  has dimension $\Delta={d-2\over 2}$, so when $d=-2$ we have $\Delta=-2$, and each term in the action \eqref{eqsgal} is separately invariant under
  \be D\phi= -x^\mu\partial_\mu\phi+2\phi \,. \label{scalesymegal}\ee
  
  As with DBI, if conformal symmetry is to fix the theory as claimed in \cite{Cheung:2020qxc},  the special conformal transformations should include non-linear parts.  The form of the special conformal generators which does this is
\be 
K^\mu\, \phi =\left(-2x^\mu x^\nu\partial_\nu+x^2 \partial^\mu +4 x^\mu\right)\phi -\alpha (\partial\phi)^2\partial^\mu\phi\,. \label{conformalffieldrepsenewge}
\ee
When $\alpha=0$, it reduces to the standard linear special conformal transformation \eqref{linconfsyme} with $\Delta=-2$.  The non-linear transformation \eqref{conformalffieldrepsenewge} is a symmetry of the galileon action \eqref{eqsgal} when $d=-2$, and it fixes the coefficients of the various galileon terms \cite{Nicolis:2008in} relative to each other into the special galileon combination. As with DBI, the transformation \eqref{conformalffieldrepsenewge} preserves the vacuum $\phi=0$, so we can still say that conformal symmetry is unbroken, even though it is non-linearly realized.  The transformations \eqref{scalesymegal}, \eqref{conformalffieldrepsenewge} along with the Poincare symmetries \eqref{pooincaresymse} satisfy the conformal algebra \eqref{conformalcommutatorese}, \eqref{poincaecomese} in any $d$, despited the non-linear $\alpha$ dependent terms in \eqref{conformalffieldrepsenewge}.

\subsection{Including shift symmetries\label{sgalshiftsec}}

The generators of the galileon shift symmetries are \cite{Hinterbichler:2015pqa}
\bea
&& C\phi = 1~,~~{B^\mu}\phi = x^\mu~,~~~ { S^{\mu\nu}}\phi = x^\mu x^\nu - \frac{1}{d}x^2\eta^{\mu\nu}-\alpha \left[\partial^\mu\phi\,\partial^\nu\phi-\frac{1}{d}(\partial\phi)^2\eta^{\mu\nu}\right]. \label{galshifts}
\eea
The shifts $C$, $B^\mu$ are zeroth and first other in the spacetime coordinate $x^\mu$, and they are symmetries of a generic galileon \cite{Nicolis:2008in}.  The symmetric traceless tensor shifts $S_{\mu\nu}$ are second order in $x^\mu$, and they are symmetries of the special galileon that fix the coefficients of the galileon terms relative to each other.

The shift symmetries \eqref{galshifts} do not close with the conformal symmetries of Section \ref{sgalconfsection} under the commutator.
We find 3 new transformations, two scalars and a vector and involving up to four powers of $x^\mu$, appearing on the right hand side of commutators before everything closes,
\bea && A^\mu\phi= x^2 x^\mu+\alpha\left[  4 \phi\partial^\mu\phi -2 \partial^\mu\phi x^\nu\partial_\nu\phi -x^\mu (\partial\phi)^2 \right] \,,\nn \\
&& T\phi= x^2-\alpha (\partial\phi)^2\,, \nn \\
&& N\phi= x^4-2\alpha\left[ x^2(\partial\phi)^2+2 x^\mu x^\nu\partial_\mu\phi\partial_\nu\phi-8 x^\mu\phi\partial_\mu\phi+8\phi^2\right]+\alpha^2 (\partial\phi)^4\,.\label{newtransgalee}
\eea
The new transformations $N,A^\mu$ are indeed new symmetries of the special galileon action \eqref{eqsgal} when $d=-2$.   (For example, we can quickly see that the kinetic term $\sim(\partial\phi)^2$ is invariant under the lowest order shift $N^{(0)}\phi\sim x^4$: $N^{(0)}\left((\partial\phi)^2)\right)\sim \partial_\mu\phi \partial^\mu( N^{(0)}\phi) \sim \partial_\mu\phi \, x^2 x^\mu$, which integrates by parts to $\sim (d+2)x^2 \phi$, which vanishes when $d=-2$.)
The transformation $T$ is not a symmetry, however, as we'll see it always appears in the commutators with a factor of $d+2$ which vanishes when $d=-2$.

The commutators of the original shift transformations \eqref{galshifts} and new transformations \eqref{newtransgalee} with the conformal generators are as follows.  They transform in the expected way as tensors under Lorentz transformations,
\bea && \left[J^{\mu\nu}, S^{\lambda\sigma}\right] = \eta^{ \mu\lambda} S^{\nu\sigma}- \eta^{ \nu\lambda  } S^{ \mu\sigma }+\eta^{  \mu\sigma} S^{ \lambda \nu}-\eta^{  \nu\sigma} S^{ \lambda \mu}\, ,\nn\\
&& \left[J^{\mu\nu},B^\rho\right] = \eta^{\rho\mu} B^\nu-\eta^{\rho\nu} B^\mu\, ,\nn \\
&&  \left[J^{\mu\nu},A^\rho\right] = \eta^{\rho\mu} A^\nu-\eta^{\rho\nu} A^\mu\, ,\nn \\
  &&  \left[J^{\mu\nu},C\right]= \left[J^{\mu\nu},T\right]= \left[J^{\mu\nu},N\right]=0\,. \label{newsymcommsce1}
\eea
They are eigenstates of the dilation operator,
\bea && \left [C,D \right] = 2C \, ,\nn \\
&& \left [B^\mu,D \right] = B^\mu \, ,\nn \\
&& \left [S^{\mu\nu},D \right] =\left [T,D \right] = 0 \, ,\nn \\
&& \left [A^\mu,D \right] = -A^\mu\, , \nn \\
&& \left [N,D \right] = -2N \, , \label{newsymcommsce2} 
\eea
and they transform into each other under $K,P$,
\bea 
&&\left[P^\mu,N\right ] =4 A^\mu\, ,\nn\\
&&\left[P^\mu,A^\nu \right ] =2 S^{\mu\nu}+{d+2\over d} \eta^{\mu\nu}T \, ,\nn\\
&&\left[P^\mu,T\right ] =2 B^\mu\, ,\nn\\
&&\left[P^{\mu}, S^{\nu \lambda}\right] = \eta^{\mu\nu}B^\lambda+\eta^{\mu\lambda}B^\nu-\frac{2}{d}B^\mu\eta^{\nu\lambda}~,\nn \\
&&\left [P^\mu,B^\nu\right ] = \eta^{\mu \nu }C \, ,\nn\\
 &&\left[P^\mu,C\right ] =0 \, , \label{newsymcommsce3}
\eea
\bea 
 &&\left[K^\mu,C\right ] =-4 B^\mu\, , \nn\\
 &&\left [K^\mu,B^\nu\right ] =-2 S^{\mu\nu}-{d+2\over d} \eta^{\mu\nu}T\, , \nn\\
 &&\left[K^{\mu}, S^{\nu \lambda}\right] =-  \eta^{\mu\nu}A^\lambda-\eta^{\mu\lambda}A^\nu+\frac{2}{d}A^\mu\eta^{\nu\lambda}~,\nn \\
 &&\left[K^\mu,T\right ] =-2 A^\mu\, ,\nn\\
&&\left[K^\mu,A^\nu \right ] =- \eta^{\mu\nu}N\, , \nn\\
&&\left[K^\mu,N\right ] =0 \,.\label{newsymcommsce4}
\eea
These relations can be summarized in this figure,
\be
\ \ \ \raisebox{-96pt}{\epsfig{file=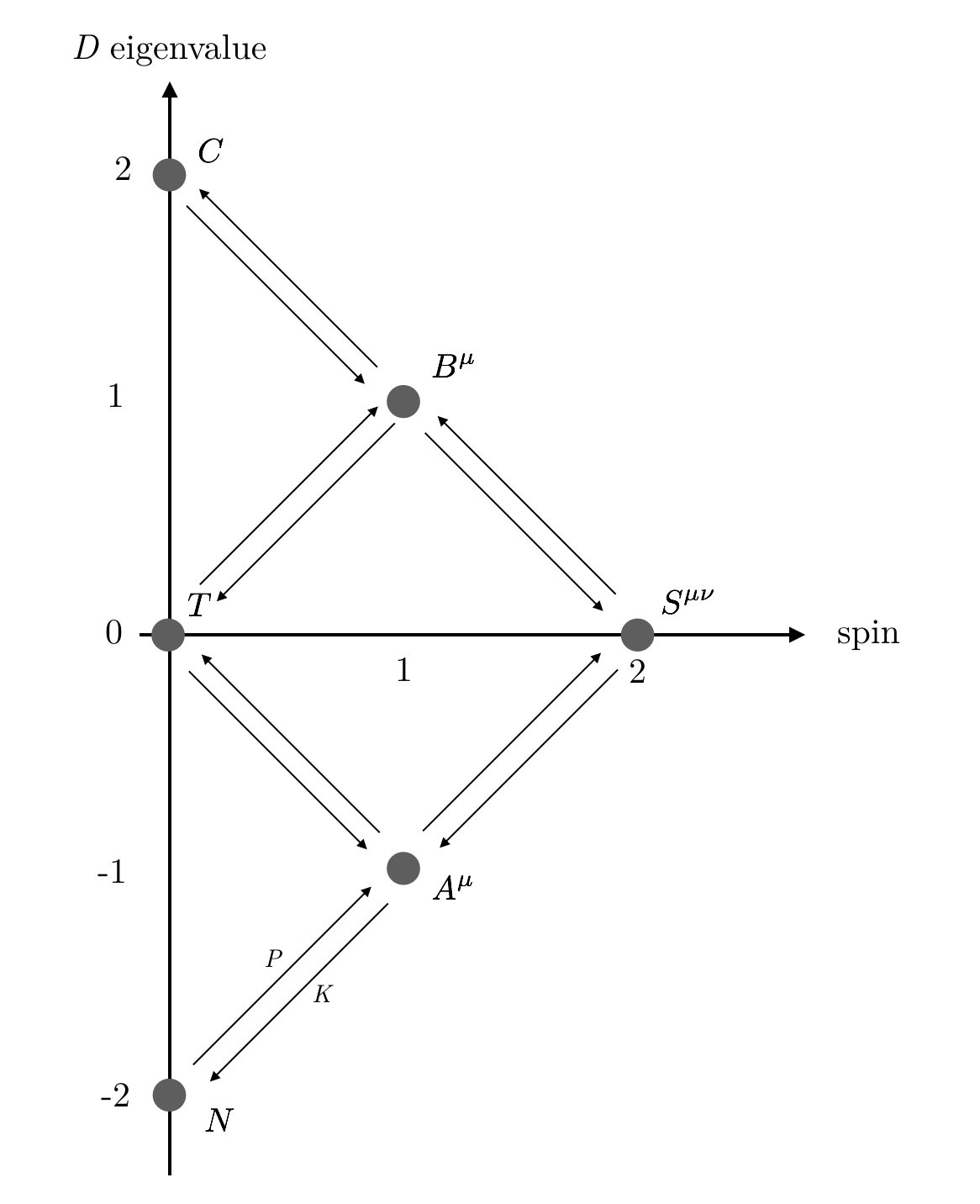,height=2.7in}}\label{sgalmodule}
\ee

The shift transformations thus form a finite dimensional Verma module of the conformal algebra where $N$ is a conformal primary of weight $-2$.  This Verma module describes a symmetric traceless tensor representation of the conformal algebra $so(2,d)$.   Group the shift generators into a $d+2$ dimensional symmetric $g$-traceless matrix, $S^{AB}=S^{BA},\ S^{AB}g_{AB}=0$, with $g$ as in \eqref{D2metricdefee}, 
\be S^{AB}= \left(\begin{array}{c|c|c}  \frac{1}{4} (C+N+2 T)  & {1\over 4}(C-N) & - {1\over 2}\left(A^\mu+B^\mu\right) \\\hline  {1\over 4}(C-N)  & \frac{1}{4} (C+N-2 T) & {1\over 2}\left(A^\mu-B^\mu\right) \\\hline - {1\over 2}\left(A^\mu+B^\mu\right) & {1\over 2}\left(A^\mu-B^\mu\right)& S^{\mu\nu} +{1\over D}\eta^{\mu\nu}T \end{array}\right).\label{Smatrixgenembsee}\ee
Grouping the conformal generators as in \eqref{Jmatrixgenembsee}, the commutators \eqref{newsymcommsce1}, \eqref{newsymcommsce2}, \eqref{newsymcommsce3}, \eqref{newsymcommsce4} become simply
\be \left[J^{AB}, S^{C D }\right] = g^{ AC } S^{BD }- g^{ BC   } S^{ AD  }+g^{  AD } S^{ C  B}-g^{  BD } S^{ C  A}\,,\label{JSsgalcommee}\ee
which is the statement that $S^{AB}$ transforms as a symmetric traceless tensor representation under the conformal algebra.

The commutators of the shift generators with each other give back conformal generators, and are all proportional to $\alpha$,
\begin{align}
\left [S^{\mu\nu },S^{\lambda\sigma }\right ] &=-\alpha\left(\eta^{\mu\lambda}J^{\nu\sigma }+\eta^{\nu\lambda }J^{\mu\sigma }+\eta^{\nu\sigma}J^{\mu\lambda }+\eta^{\mu\sigma}J^{\nu\lambda }\right), \nn\\ \nn
\left[B^{\mu}, S^{\nu \lambda}\right] & =\alpha\Big( \eta^{\mu\nu}P^\lambda+\eta^{\mu\lambda}P^\nu-\frac{2}{d}P^\mu\eta^{\nu\lambda}\Big),  \\ \nonumber
\left[A^{\mu}, S^{\nu \lambda}\right] & =- \alpha\Big( \eta^{\mu\nu}K^\lambda+\eta^{\mu\lambda}K^\nu-\frac{2}{d}K^\mu\eta^{\nu\lambda}\Big),  \\ \nonumber
\left [S^{\mu\nu },C\right ] &=\left [S^{\mu\nu },T\right ]=\left [S^{\mu\nu },N\right ]=0\,,\nn\\
\left[A^{\mu}, A^\mu \right] & =\left[B^{\mu}, B^\nu\right] =0\,,  \nn\\
\left[A^{\mu}, B^\mu \right] & =-2\alpha\left(\eta^{\mu\nu}D-J^{\mu\nu}\right)\,, \nn\\
\left[A^{\mu}, T \right] & =-2\alpha K^\mu,\ \ \ \left[A^{\mu}, C \right]  =4\alpha P^\mu, \ \ \left[A^{\mu}, N \right]  =0,\nn\\
\left[B^{\mu}, T \right] & =2\alpha P^\mu,\ \ \ \left[B^{\mu}, N \right]  =-4\alpha K^\mu, \ \ \left[B^{\mu}, C \right]  =0,\nn\\
\left[N,C\right] &=16\alpha D,\ \ \left[N,T\right] =\left[C,T\right] =0. \label{shitshiftcgale}
\end{align}
In terms of \eqref{Smatrixgenembsee} and \eqref{Jmatrixgenembsee}, the commutators \eqref{shitshiftcgale} become simply
\be \left[S^{A B },S^{C D  }\right ] =-\alpha\left(g^{A C }J^{BD  }+g^{BC  }J^{A D  }+g^{BD }J^{A C  }+g^{A D }J^{BC  }\right)\,.\label{SSsgalcomme} \ee

We can now identify the full algebra.  Put together the anti-symmetric conformal generators \eqref{Jmatrixgenembsee} and the symmetric traceless shift generators \eqref{Smatrixgenembsee} into a traceless matrix as follows
\be M^{AB} \equiv -{1\over 2}\left(J^{AB}+{1\over \sqrt{\alpha}}S^{AB}\right)\,, \ \ \alpha>0 .\label{specfgaealphge0e}\ee
Then the commutators \eqref{SSsgalcomme}, \eqref{JSsgalcommee}, \eqref{Jcommsd2e} can be summarized as
\be \left[ M^{AB},M^{CD}\right]=g^{CB}M^{AD}-g^{AD}M^{CB}\,,\ee
which are the commutation relations of $sl(d+2,{\mathbb R})$.
The smallest symmetry algebra of the special galileon that includes the conformal symmetry of \cite{Cheung:2020qxc} and all the extended shift symmetries is thus isomorphic to $sl(d+2,{\mathbb R})$.  In writing \eqref{specfgaealphge0e} we have assumed $\alpha>0$, the ``right sign'' from the point of view of positivity bounds \cite{deRham:2017imi,Roest:2020vny}.  If $\alpha<0$, then the algebra would instead be the maximally split real form.

The breaking pattern is to the $d$ dimensional conformal algebra, since the shifts $S^{AB}$ are the only ones that do not preserve the vacuum $\phi=0$,
\be sl(d+2,{\mathbb R})\rightarrow so(2,d)\, ,\ \ \ \alpha>0.\ee
In fact, this is the same symmetry breaking pattern as the AdS special galileon theory in $d+1$ dimensions (the $k=2$ non-abelian AdS scalar shift theory of \cite{Bonifacio:2018zex}). 

\section{Weyl symmetry\label{DBIweylsec}}

A conformal field theory can generally be coupled to a background metric $g_{\mu\nu}$ in such a way that the conformal symmetry descends from a local Weyl symmetry \cite{Farnsworth:2017tbz,Wu:2017epd} (there are some exceptions in non-unitary theories, see e.g. \cite{Karananas:2015ioa,Brust:2016gjy,Nakayama:2019xzz}).
The usual form of the Weyl symmetry for a field $\phi$ of weight $\Delta$ is
\be \delta g_{\mu\nu}=2\sigma g_{\mu\nu},\ \ \ \delta\phi=-\Delta\sigma\phi,\label{weylstande}\ee
where $\sigma(x)$ is the local scalar gauge parameter.
This symmetry is present in addition to ordinary diffeomorphism (diff) symmetry with local parameter $\xi^\mu(x)$,
\be \delta g_{\mu\nu}=\nabla_\mu\xi_\nu+\nabla_\nu\xi_\mu,\ \ \ \delta\phi=\xi^\mu\partial_\mu\phi.\label{weylstande2}\ee

The global conformal transformations emerge when we go to flat space, $g_{\mu\nu}=\eta_{\mu\nu}$, and restrict to those diff+Weyl symmetries which preserve the flat metric $\eta_{\mu\nu}$.   Such symmetries are those for which $\xi^\mu$ is a conformal Killing vector, $\partial_\mu\xi_\nu+\partial_\nu\xi_\nu={2\over d}\partial^\lambda  \xi_\lambda \eta_{\mu\nu}$,
 and $\sigma=-{1\over d}\partial^\mu  \xi_\mu$.  The conformal Killing vector corresponding to dilations is $\xi^\mu=-x^\mu$ with $\sigma=1$.  Plugging this into the scalar transformations in \eqref{weylstande}, \eqref{weylstande2} gives the standard dilation transformation $\delta\phi= -\left(x^\mu\partial_\mu+\Delta\right)\phi.$
 The conformal Killing vector corresponding to special conformal transformations is $\xi^\mu=-2b \cdot x\, x^\mu+x^2 b^\mu $ with $\sigma=2 b\cdot x$, where $b^\mu$ is a constant vector parametrizing the transformations.  Plugging this into the scalar transformations in \eqref{weylstande}, \eqref{weylstande2} gives the standard linear special conformal transformation $\delta\phi=b_\mu\left(-2x^\mu x^\nu\partial_\nu+x^2 \partial^\mu -2 x^\mu\Delta\right)\phi$.
 
 \subsection{DBI}

To reproduce the flat space non-linear terms in \eqref{conformalffieldrepsenew} after coupling to a general background metric, we need new non-linear terms proportional to $\alpha$ in the Weyl transformation.   We will assume that the Weyl transformation for the metric, and the diffeomorphisms, retain their standard forms and do not depend on $\alpha$.  The Weyl transformation for the scalar will be modified.  Each power of $\alpha$ comes with 2 derivatives and 2 powers of $\phi$, so the most general possible modification at order $\alpha$ that reduces to \eqref{scalesyme} for $\sigma=1$ and \eqref{conformalffieldrepsenew}  for $\sigma=2 b\cdot x$ is
\be \delta\phi=\sigma\phi+\alpha\left( {1\over 2}  \phi^2\partial_\mu\phi \partial^\mu \sigma+a_1 \phi^3 R\sigma+a_2 \phi^3\square\sigma \right)+{\cal O}\left(\alpha^2\right)\, ,\label{possibleweyl}\ee
with constants $a_1,a_2$.

We now want to couple \eqref{lageeqflat} to the metric such that this new Weyl transformation is a symmetry when $d=0$.  The most general coupling to the metric, to order $\alpha$, is
\bea && S=\int d^dx\  \sqrt{-g}\bigg[{1\over \alpha}-{1\over \alpha }\sqrt{1+\alpha (\partial\phi)^2}+b_0 R \phi^2\\
&&+\alpha\left( b_1 R^{\mu\nu}\phi^2\partial_\mu\phi\partial_\nu\phi+ b_2 R \phi^2(\partial\phi)^2 +b_3 R\phi^3\square\phi+ b_4 W_{\mu\nu\lambda\rho}^2 \phi^4+ b_5  R_{\mu\nu}^2\phi^4+ b_6 R^2\phi^4 \right) +{\cal O}\left(\alpha^2\right)\bigg]\, , \nn\label{lageeqcurved}\eea
where $W_{\mu\nu\lambda\rho}$ is the Weyl tensor and $b_0,\ldots, b_6$ are constants.  Requiring Weyl invariance up to order $\alpha$ when $d=0$ then fixes $a_1=0$, $b_0 = -1/4$, $b_1=1/4$, $b_2=0$, $b_3=-a_2/2$,  $b_5=1/32$, $b_6=a_2/4$, while leaving $a_2,b_4$ free.

Thus the Weyl symmetry and invariant action to order $\alpha$ is
\bea &&  S=\int d^dx\  \sqrt{-g}\bigg[{1\over \alpha}-{1\over \alpha }\sqrt{1+\alpha (\partial\phi)^2} -{1\over 4} R\phi^2 \nn\\
&&+\alpha\left({1\over 4} R^{\mu\nu}\phi^2\partial_\mu\phi\partial_\nu\phi+ {1\over 32} R_{\mu\nu}^2\phi^4 -{a\over 2} R\phi^3\square\phi+ {a\over 4} R^2\phi^4+ b W_{\mu\nu\lambda\rho}^2 \phi^4\right)  +{\cal O}\left(\alpha^2\right) \bigg]\,, \nn\\
&&  \delta g_{\mu\nu}=2\sigma g_{\mu\nu},\ \ \ \  \delta\phi=\sigma\phi+\alpha\left( {1\over 2}  \phi^2\partial_\mu\phi \partial^\mu \sigma+a \phi^3\square\sigma \right)+{\cal O}\left(\alpha^2\right)\,. \label{weylinvacte}
\eea
where $a$, $b$ are the two unfixed free parameters. The parameter $a$ can be further removed by the field redefinition 
\be \phi \rightarrow \phi +\tfrac{1}{2} a \alpha R\phi^3\, ,\label{fieldrredefwelyape}\ee
as in \cite{Farnsworth:2017tbz}.  This is the only allowed field redefinition at order $\alpha$ that does not transform the scalar, and hence does not modify the DBI form of the Lagrangian, when restricted to flat space.
The parameter $b$ remains arbitrary since it comes in front of the Weyl-squared term, which is separately Weyl invariant to lowest order. This term would presumably be completed at higher order into some invariant of the $\alpha$-deformed Weyl symmetry, or else forced to vanish if no such invariant exists.  Note that the coupling to the metric generically breaks all the shift symmetries. 

This new Weyl symmetry satisfies the same algebra as the standard Weyl symmetry, namely that the Weyl transformations are abelian,
\be \left[\delta_{\sigma_1},\delta_{\sigma_2}\right]=0,\label{weyldbicomee}\ee
 up to order $\alpha$.  In fact this requirement alone is enough to fix the form of the transformation: demanding \eqref{weyldbicomee} alone for \eqref{possibleweyl} when $d=0$ fixes $a_1=0$, and then taking into account the possible field re-definition \eqref{fieldrredefwelyape} removes $a_2$.
 
 In \cite{Cheung:2020qxc}, an action linear in the curvatures and including all powers of $\phi$ was provided which makes the stress tensor on flat space traceless.  The terms linear in the curvature in the action \eqref{weylinvacte} match the terms linear in the curvature from \cite{Cheung:2020qxc}, up to order $\alpha$ (and when $a=0$).  However the action in \cite{Cheung:2020qxc} cannot itself be Weyl invariant in the manner we are requiring, because, as we can see from \eqref{weylinvacte}, terms quadratic in the curvature are required. 
 
Pushing to next order in $\alpha$ we find that the action can be made Weyl invariant when $d=0$ to quadratic order in $\alpha$, with the addition of the following terms to the action,
\begin{align}
 S\bigg|_{\alpha^2}=&\ \int d^dx\  \sqrt{-g}\, \bigg\{  -\frac{\alpha^2}{32} R\phi^2 (\partial\phi)^4-\frac{\alpha^2}{8} R^{\mu\nu} \phi^2 (\partial\phi)^2 \nabla_\mu \phi \nabla_\nu \phi \label{orderalpha2actione} \\
&+\frac{\alpha^2}{64}\phi^4\bigg[ \left(R_{\mu\nu}^2 +\tfrac{1}{2}R^2\right) (\partial\phi)^2-2R R^{\mu\nu} \nabla_\mu \phi \nabla_\nu \phi-6 R^{\mu\nu} R_{\nu\sigma}  \nabla_\mu \phi \nabla^\sigma \phi \bigg]\nonumber\\
& +\frac{\alpha^2}{48}W_{\mu\nu\rho\sigma} \bigg[2\phi^3   \nabla^\mu \phi \nabla^\rho \phi \nabla^\nu \nabla^\sigma \phi  -3R^{\mu\rho} \phi^4 \nabla^\nu \phi \nabla^\sigma \phi -\tfrac{1}{6} \phi^6\nabla^\sigma \nabla^\nu  R^{\mu\rho}  -\phi^5 \nabla^\nu  R^{\mu\rho} \nabla^\sigma\phi  \bigg]\nonumber\\
& -\frac{\alpha^2}{192} \phi^6\bigg[W_{\mu\nu\rho\sigma} R^{\mu\rho} R^{\nu\sigma}+R^{\mu\nu} R_{\nu\tau} R^\tau_\mu-\tfrac{1}{4} R^3  \bigg]  \bigg\}\, ,\nonumber
\end{align}
along with the additional non-linear terms in the Weyl transformation,
\begin{align}
\delta \phi\bigg|_{\alpha^2} = \frac{\alpha^2}{4} S^{\mu\nu}\phi^4 \nabla_\mu \phi \nabla_\nu \sigma. \label{orderalpha2weyle}
\end{align}
Here $S_{\mu\nu}$ is the Schouten tensor, defined in general $d$ as
\begin{align}
S_{\mu\nu} = \frac{1}{(d-2)} \left(R_{\mu\nu} - \frac{1}{2(d-1)} R g_{\mu\nu}\right).
\end{align}
This action and Weyl transformation will also have ambiguities due to field re-definitions that vanish in flat space (coming from order $\alpha^2$ terms in \eqref{fieldrredefwelyape}), as well as ambiguities from possible Weyl invariants.  Here we have not attempted to parameterize all these ambiguities but have simply made a choice for which the action and Weyl transformation are manageable, in order to illustrate that there is no obstruction to extending the Weyl symmetry to order $\alpha^2$.  
The first line of \eqref{orderalpha2actione} matches those found in the action linear in the curvatures constructed in \cite{Cheung:2020qxc} by demanding the theory have a vanishing stress tensor in flat space.   In the third line of \eqref{orderalpha2actione} there are additional terms linear in the curvature that do not appear in \cite{Cheung:2020qxc}, but these are proportional to the Weyl tensor and so they do not contribute to the flat space stress tensor. 
Thus what we find is consistent with \cite{Cheung:2020qxc} up to order $\alpha^2$.  It would be interesting if the action in \cite{Cheung:2020qxc} could be extended to an action which is fully Weyl invariant to all orders.

\subsection{Special Galileon}
A similar procedure can be applied in the case of the special galileon in $d = -2$, and once again we find that the theory can be improved to be Weyl invariant, at least to $\mathcal{O}(\alpha)$. The non-linear conformal transformations \eqref{scalesymegal},  \eqref{conformalffieldrepsenewge} get uplifted to 
\begin{align}
\delta \phi  = 2 \sigma \phi - \frac{\alpha}{2} (\partial\phi)^2 \nabla_\mu \sigma \nabla^\mu \phi \,,\label{weyltanseqe}
\end{align}
in a general background metric.  A priori there are many more terms that could enter this transformation, however requiring that the Weyl transformations commute as in \eqref{weyldbicomee} reduces this number, and employing field redefinitions that vanish on flat space analogous to \eqref{fieldrredefwelyape} removes the rest.
We find the following action is Weyl invariant under \eqref{weyltanseqe} in $d=-2$, up to order $\alpha$,
\begin{align}
 S =& \int d^dx \sqrt{-g}\, \bigg\{ -\frac{1}{2}(\partial\phi)^2 -\frac{1}{6} R \phi^2- \frac{\alpha}{12}(\partial\phi)^2\Big[(\square\phi)^2 - (\partial_\mu\partial_\nu\phi)^2\Big] \\
 &   -\frac{\alpha }{12} R^{\mu\nu} \phi^2 \nabla_\mu \nabla^\tau \phi \nabla_\nu \nabla_\tau \phi +\frac{\alpha}{72} R \phi^2 \nabla_\mu \nabla_\nu \phi \nabla^\mu \nabla^\nu \phi \nn\\
 &+ \frac{\alpha}{12} R^{\mu\nu}\phi^2 \Box \phi   \nabla_\mu \nabla_\nu \phi -\frac{ \alpha}{72} R \phi^2(\Box \phi)^2\nn\\
&-\frac{\alpha }{35} W^{\mu\nu\rho\sigma}\phi  \bigg[\frac{13}{6}  \nabla_\rho \phi\nabla_\mu \phi\nabla_\sigma\nabla_\nu \phi -\phi\nabla_\rho \nabla_\mu \phi\nabla_\sigma\nabla_\nu \phi+W_{\tau\rho\nu\sigma}\phi\nabla_\mu \phi \nabla^\tau \phi\bigg]\nonumber\\
&-\frac{\alpha }{48}\bigg[9 R^{\mu\nu} R_{\mu\tau}\phi^2\nabla^\tau\phi \nabla_\nu \phi -\frac{7}{2}R_{\mu\nu}^2 (\partial\phi)^2 \phi^2-\frac{1}{6} R^2 (\partial\phi)^2 \phi^2-3R R^{\mu\nu} \phi^2 \nabla_\mu \phi \nabla_\nu \phi\bigg]\nonumber\\
&-\frac{ \alpha}{210} \bigg[15  R^{\nu\rho}\nabla^\sigma W_{\mu\nu\rho\sigma}\phi^3\nabla^\mu  \phi-14  W_{\mu \nu  \rho \sigma} R^{\mu \rho}  \phi^2\nabla^\sigma \phi \nabla^\nu \phi-\frac{1}{2}  R^{\mu\rho}\nabla^\nu  \nabla^\sigma W_{\mu\nu\rho\sigma}  \phi^4\bigg]\nonumber\\
&+\frac{\alpha}{48}\bigg[\frac{1}{7} W_{\mu\nu\rho\sigma} R^{\mu\rho} R^{\nu\sigma} -\frac{1}{2}R_{\mu\nu}R^{\mu\rho} R_{\rho}^\nu+\frac{1}{4} R_{\mu\nu}^2 R +\frac{1}{36}R^3\bigg]\phi^4 \bigg\} +{\cal O}\left(\alpha^2\right)\,.\nonumber
\end{align}
The terms in the second line are the same as those in \cite{Cheung:2020qxc} found by demanding tracelessness of the stress tensor in flat space\footnote{Our expressions match those in \cite{Cheung:2020qxc} for $\alpha = -1$.}.  In the third line there are additional terms linear in the curvature that do not appear in \cite{Cheung:2020qxc}, but these can be removed by field re-definitions that vanish in flat space with $\sigma = 2b\cdot x$, at the expense of making the Weyl transformation \eqref{weyltanseqe} more complicated.   
Although admittedly not very illuminating, we present these results both to demonstrate there is no immediate obstruction to Weyl invariance, and in the hope that these actions may be found to arise from a more fundamental understanding of the Weyl invariance of these theories to all orders in $\alpha$.

\section{Conclusions}

We have found the field transformations behind the conformal symmetries for DBI in dimension $d=0$ and the special galileon in dimension $d=-2$ that were found in \cite{Cheung:2020qxc}.  Using these, we have studied how these symmetries interplay with the extended shift symmetries, finding new symmetries that close them into a larger algebra.  In the DBI case, this larger algebra can be seen as the conformal algebra of the $d+1$ dimensional target space in the embedded brane way of understanding DBI.  In the special galileon case it is a real form of the special linear algebra (it would be interesting if this could be understood from a geometric picture of the special galileon \cite{Novotny:2016jkh,Roest:2020vny}).
We have also found the corresponding Weyl invariances, to second order in the coupling in the DBI case and to lowest order in the coupling in the special galileon case.

A natural question is whether theories with higher order shift symmetries \cite{Hinterbichler:2014cwa,Griffin:2014bta}, $\delta\phi\sim x^k$ for $k>2$, could emerge and have conformal symmetry.  Continuing the pattern we see here for DBI ($k=1$) and special galileon $(k=2)$, we would expect such a theory to be conformal in dimension $d=-2(k-1)$, where the scalar has weight $\Delta=-k$.  We would expect the shift symmetries and new symmetries found by commuting with the conformal generators to form a rank $k$ symmetric-traceless representation of the $so(2,d)$ conformal algebra, where the conformal primary has weight $-k$.  However, as shown in \cite{Bonifacio:2018zex}, the commutators of these generators among themselves cannot close back to the conformal algebra when $k>2$, except in the trivial case where such commutators are all abelian, which would presumably correspond to a free theory or one whose structure is not fixed in a non-trivial way by the symmetry (this is consistent with the results of \cite{Bogers:2018zeg,Roest:2019oiw} forbidding a non-trivial algebra for $k>0$ theories in flat space).  This conclusion does not apply if multiple fields of higher spin are allowed, for which non-trivial algebras are possible \cite{Joung:2015jza,Bonifacio:2018zex,Bonifacio:2019hrj}.

Given that these theories have a formal conformal invariance in unphysical $d$, it would be interesting to explore whether the tools of the modern conformal bootstrap \cite{Rattazzi:2008pe,Poland:2018epd} could be extended to these unphysical dimensions and brought to bear in order to non-perturbatively explore these theories.  An immediate obstacle is that the conformal transformations we have are non-linear and so the fields do not transform as conformal primaries in the usual way, so the form of correlators, crossing symmetry, etc. are all presumably modified.  Alternatively, perhaps there are composite operators that transform in the usual linear way as conformal primaries, given that the basic field $\phi$ transforms non-linearly.   If such operators exist, their correlators could be a target for the bootstrap.  But there is no guarantee that they should exist, especially given expectations that these soft scalar theories are not true local field theories with local operators all the way to the UV \cite{Dubovsky:2012wk,Keltner:2015xda}.

\vspace{-.2cm}
\paragraph{Acknowledgments:} We would like to thank James Bonifacio, Austin Joyce and David Stefanyszyn for helpful conversations. KH acknowledges support from DOE grant DE-SC0019143, KH and KF acknowledge support from Simons Foundation Award Number 658908. OH was supported by the GA\v{C}R Grant EXPRO 20-25775X.   


\renewcommand{\em}{}
\bibliographystyle{utphys}
\addcontentsline{toc}{section}{References}
\bibliography{DBIconformalarxiv}

\end{document}